# MICROBURST WINDSPEED POTENTIAL ASSESSMENT: PROGRESS AND RECENT DEVELOPMENTS

Kenneth L. Pryor
Center for Satellite Applications and Research (NOAA/NESDIS)
Camp Springs, MD

## 1. Introduction

Recent testing and validation have found that the Geostationary Operational Environmental Satellite (GOES) microburst products are effective in the assessment and short-term forecasting of downburst potential and associated wind gust magnitude. Two products, the GOES sounder Microburst Windspeed Potential Index (MWPI) and the multispectral GOES imager product have demonstrated capability in downburst potential assessment (Pryor 2009). Both the GOES sounder MWPI and imager microburst risk products are predictive linear models developed in the manner exemplified in Caracena and Flueck (1988). Each microburst product consists of a set of predictor variables that generates output of expected microburst risk. This paper compares and contrasts the sounder and imager microburst products and outlines the advantages of each product in the nowcasting process. An updated assessment of the sounder MWPI and imager microburst products, case studies demonstrating effective operational use of the microburst products, and validation results for the 2009 convective season over United States Great Plains is presented.

## 2. Methodology

### 2.1 *GOES Imager Microburst Product*

The objective of this validation effort was to qualitatively and quantitatively assess the performance of the GOES imager-derived microburst product by employing classical statistical analysis of real-time data. Wakimoto (1985) and Atkins and Wakimoto (1991) discussed the effectiveness of using mesonetwork surface observations and radar reflectivity data in the verification of the occurrence of downbursts. Well-defined peaks in wind speed (Wakimoto 1985; Atkins and Wakimoto 1991) were effective indicators of downburst occurrence. It was found that derived product imagery generated one to three hours prior to downburst events provided an optimal characterization of the pre-convective thermodynamic environment over the southern Great Plains and Great Basin.

Product images were generated by Man computer Interactive Data Access System (McIDAS)-X where GOES imager data was read and processed, and brightness temperature differences were calculated and visualized with a color enhancement in which moderate to high microburst risk was indicated as a progression from yellow to red shading. The image data consisted of derived brightness temperatures from infrared bands 3, 4, and 5, obtained from the Comprehensive Large Array-data Stewardship System (CLASS, http://www.class.ncdc.noaa.gov/) for archived data and the McIDAS Abstract Data Distribution Environment (ADDE) server for real-time data. Microburst algorithm output was also visualized by McIDAS-V software (version 1.0beta4) (Available online at

http://www.ssec.wisc.edu/mcidas/software/v/). A contrast stretch and built-in color enhancement were applied to the output images to highlight regions of elevated microburst risk. In addition, the three-band microburst risk algorithm was implemented in McIDAS-V, where, for selected events, the program was utilized to retrieve real-time GOES imager data and produce a color-enhanced product image. Visualizing algorithm output in McIDAS-V allowed for cursor interrogation of output brightness temperature and more precise recording of BTD values associated with observed downburst events. This methodology has served to increase the statistical significance of the relationship between output BTD and microburst wind gust magnitude.

Transmittance weighting functions from bands 3, 4, and 5 that specify the relative contribution from each atmospheric layer to emitted radiation were obtained over Amarillo, Texas from a Cooperative Institute for Meteorological Satellite Studies (CIMSS) website (http://cimss.ssec.wisc.edu/goes/wf/index.php). The weighting functions were compared to temperature profiles in the 0000 UTC Amarillo, Texas RAOB and product imagery for selected microburst events that occurred during the evening hours (after 0000 UTC or 1800 LST). The site of Amarillo was selected due to its location in the southern High Plains and proximity to mesonet stations in western Oklahoma and Texas. Figure 1 displays GOES-11 imager and sounder transmittance weighting functions over Amarillo, Texas. The black curve represents the weighting function for imager band 3 (6.7µm) while the tan and brown curves represent functions for bands 4 (11µm) and 5 (12 µm), respectively. The imager and sounder bands 3 and 11 (7µm) weighting function peaks were located near the 400-mb and 500-mb levels, respectively, corresponding roughly to the mid-tropospheric moist layer (moist adiabatic lapse rate layer) displayed in the RAOB. The band 3 and sounder band 11 weighting function peaks, overlying peaks in bands 4 and 5 functions at the surface, correlated well with the unstable sounding profile that was dry adiabatic below 500 mb with a superadiabatic layer near the surface. The combined display of the temperature curve of the RAOB and weighting functions indicated that the mid-tropospheric moist layer was vertically extensive and was overlying a dry, convective boundary layer, thus, illustrating favorable environmental conditions for strong convective downdrafts. The microburst risk (output BTD of 46K) calculated over Amarillo using brightness temperatures associated with the GOES-11 weighting functions was very close to the GOES BTD of 47 near Lubbock at the time a severe downburst was in progress. Thus, the group of spectral bands selected to detect radiation emitted from layers of interest in the atmosphere could be determined to be most effective for the purpose of inferring the presence of a favorable environment for microbursts.

For each microburst event, product images were compared to radar reflectivity imagery and surface observations of convective wind gusts as provided by Oklahoma and West Texas Mesonet stations (Brock et al. 1995; Schroeder et al. 2005). A detailed description of the validation process using INL Mesonet data is available in Pryor (2009). Next Generation Radar (NEXRAD) base reflectivity imagery (levels II and III) from National Climatic Data Center (NCDC), and University Corporation for Atmospheric Research (UCAR) Internet Data Distribution (IDD) server was utilized to verify that observed wind gusts were associated with downbursts and not associated with other types of convective wind phenomena (i.e. gust fronts). Another application of the NEXRAD imagery was to infer microscale physical properties of downburst-producing convective storms. Particular radar reflectivity signatures, such as the rear-inflow notch (RIN)(Przybylinski 1995) and the spearhead echo (Fujita and Byers 1977), were effective indicators of the occurrence of downbursts.

Covariance between the variables of interest, microburst risk (expressed as output brightness temperature) and surface downburst wind gust speed, was analyzed to assess the performance of the imager microburst algorithm. A very effective means to quantify the functional relationship between microburst index algorithm output and downburst wind gust strength at the surface was to calculate correlation between these variables. Thus, correlation between GOES imager microburst risk and observed surface wind gust velocities for the selected events were computed to assess the significance of this functional relationship. Statistical significance testing was conducted, in the manner described in Pryor and Ellrod (2004), to determine the confidence level of correlations between observed downburst wind gust magnitude and microburst risk values. The confidence level is intended to quantify the robustness of the correlation between microburst risk values and wind gust magnitude.

## 2.2 GOES Sounder Microburst Product

In a similar manner to the imager product, this validation effort assessed and intercompared the performance of the GOES sounder-derived microburst products by employing classical statistical analysis of real-time data. Accordingly, this effort entailed a study of downburst events over the southern Great Plains region. Again, this validation was executed in a manner that emulated historic field projects such as the 1982 Joint Airport Weather Studies (JAWS) (Wakimoto 1985) and the 1986 Microburst and Severe Thunderstorm (MIST) project (Atkins and Wakimoto 1991). Data from the GOES MWPI product was collected over Oklahoma and western Texas for downburst events that occurred between 1 June 2007 and 30 September 2009. Microburst index values were then validated against surface observations of convective wind gusts as recorded by Oklahoma and West Texas Mesonet stations. Wakimoto (1985) and Atkins and Wakimoto (1991) discussed the effectiveness of using mesonet surface observations and radar reflectivity data in the verification of the occurrence of downbursts. Since surface data quality is paramount in an effective validation program, Oklahoma and western Texas were chosen as study regions due to the wealth of high quality surface observation data provided by the Oklahoma and West Texas Mesonets, a thermodynamic environment typical of the Great Plains regions during the warm season, and relatively homogeneous topography. Pryor (2008a) discussed the importance of the dryline (Schafer 1986) in convective storm climatology in the Southern Plains region as well as the selection of the High Plains region for a validation study.

Downburst wind gusts, as recorded by mesonet observation stations, were measured at a height of 10 meters (33 feet) above ground level. In order to assess the predictive value of GOES microburst products, the closest representative index values used in validation were obtained for retrieval times one to three hours prior to the observed surface wind gusts. Representativeness of proximate index values was ensured by determining from analysis of surface observations, and radar and satellite imagery, that no change in environmental static stability and air mass characteristics between product valid time and time of observed downbursts had occurred. Previous field studies of microbursts, especially the MIST project in northern Alabama (Atkins and Wakimoto 1991), noted that peak wind gusts are typically recorded three to five kilometers from the point of impact. Beyond a distance of five kilometers, interaction of the convective storm outflow with the surface, most likely through the process of friction, will result in a decrease in downburst wind velocity and subsequently, wind gust measurements that are unrepresentative of the ambient thermodynamic environment. Ideally, at the time of downburst occurrence, the distance between the parent storm cell and the location of the measured downburst wind gust should be no more than five kilometers. An additional criterion for inclusion into the data set was a wind gust measurement of at least F0 intensity (35 knots) on the Fujita scale (Fujita 1971). Wind gusts of 35 knots or greater are considered to be operationally significant for transportation, especially boating and aviation. An algorithm devised by Wakimoto (1985) to visually inspect wind speed observations over the time intervals encompassing candidate downburst events was implemented to exclude gust front events from the validation data set. In summary, the screening process employed to build the validation data set that consisted of criteria based on surface weather observations and radar reflectivity data yielded a sample size of 168 downbursts and associated index values.

Similar to the validation process for the imager microburst product, for each microburst event, product images were compared to radar reflectivity imagery. Again, Next Generation Radar (NEXRAD) base reflectivity imagery was utilized for this purpose. Covariance between the variables of interest, MWPI and surface downburst wind gust speed, was considered. Algorithm effectiveness was assessed as the correlation between MWPI values and observed surface wind gust velocities. Statistical significance testing was conducted to determine the confidence level of the correlation between observed downburst wind gust magnitude and microburst risk values.

## 3. Case Studies

### 3.1 *10 June 2009 High Plains Downbursts*

During the afternoon of 10 June 2009, an upper-level disturbance triggered the development of isolated convective storms over the High Plains of New Mexico, Oklahoma, and Texas in the wake of a large mesoscale convective system (MCS). Downburst wind gusts between 38 and 53 knots were recorded over the Oklahoma and Texas Panhandles and eastern New Mexico during the two-hour period, 2155 to 2355 UTC. The GOES imager microburst product indicated elevated risk values (output brightness temperature difference (BTD) >40K) in proximity to the location of the downbursts one to three hours prior to each event. In general, the pre-convective environment over the southern High Plains was favorable for hybrid microbursts, characterized by a deep convective mixed layer with a steep temperature lapse rate, especially below the 600-mb level, as indicated in the Rapid Update Cycle (RUC) model analysis soundings in Figures 3 and 4.

By 2000 UTC, near the time of maximum surface heating, the atmosphere in the wake of the MCS was destabilizing over the Texas Panhandle as evidenced by the development of enhanced cumulus clouds shown in Figure 2. Areas of enhanced cumulus are apparent along and east of an upper-level trough over eastern New Mexico and the central Texas Panhandle. Also apparent is developing convective storm activity over New Mexico that would track east through the Texas Panhandle after 2100 UTC. Figure 3 shows a GOES imager microburst product at 2000 UTC 10 June 2009 (top), with mesonet observations of downburst wind gusts plotted on the image, and a corresponding RUC sounding at Hereford, Texas. The GOES imager microburst product displayed scattered cumulus cloud (black shading) development in the unstable air mass over western Texas near the New Mexico border. In addition, isolated convective storms were developing over eastern New Mexico near the upper-level trough axis. Moderate to high microburst risk, indicated by yellow to orange shading, was in place over western Texas, near Hereford, where strong downbursts would occur about two hours later as convective storm activity tracked eastward. The RUC model sounding profile over Hereford exhibited a classic "inverted-V" and indicated that the environment over the Panhandle region was favorable for hybrid microbursts with a steep temperature lapse rate and well-developed mixed layer below 600 mb. Near 2200 UTC, downburst wind gusts of 38 and 50 knots were recorded at Dimmitt and Hereford (West Texas) mesonet stations, respectively. Note that the stronger downburst recorded at Hereford was associated with higher output BTD near 50K (orange shading).

Figure 4 displays a GOES imager microburst product at 2200 UTC 10 June (top), with mesonet observations of downburst wind gusts plotted on the image, and a corresponding RUC sounding at Kenton, Oklahoma. The GOES imager microburst product displayed convective storm activity in progress over the Texas Panhandle and eastern New Mexico, including downburst-producing storms near Hereford and Dimmitt. The RUC sounding profile over Kenton exhibited a more well-defined "inverted-V" profile. The product image and sounding profile signified that the environment over the Panhandle region was especially favorable for hybrid microbursts. Downburst wind gusts of 40 and 48 knots occurred between 2300 and 0000 UTC 11 June at Boise City and Kenton, Oklahoma mesonet stations, respectively, where a moderate to high risk of microbursts was indicated at 2200 UTC. In a similar manner to the 2000 UTC product image, this image indicated a local maximum in output BTD near the location of the downbursts that occurred as the upper-level disturbance moved over the western Oklahoma Panhandle.

Figure 5 shows GOES microburst products with overlying radar reflectivity imagery from Amarillo NEXRAD. These images, visualized by McIDAS-V software, display that downbursts were associated with multicell storms that occurred in regions of elevated microburst risk (orange shading). The combination of high radar reflectivity (> 40 dBZ) and steep sub-cloud lapse rates, as indicated in the soundings, signified that precipitation loading and sub-cloud evaporation of precipitation were both major factors in convective downdraft development, acceleration, and eventual downburst generation. For this High Plains convective storm event, the GOES imager microburst product effectively indicated downburst potential with a correlation between output BTD and wind gust speed of **.64** and a mean difference between output BTD and wind gust speed of **1.2**.

### 3.2 *26 August 2009 Cold Front Downbursts*

During the afternoon of 26 August 2009, strong convective storms developed along a weak, slow-moving cold front as it was tracking eastward over Oklahoma. Although there was very little temperature contrast across the front, the front acted as a convergence zone and a trigger for deep, moist convection. The pre-convective environment downstream of the cold front over western Oklahoma was dominated by vertical mixing that fostered the development and evolution of a convective boundary layer. Elevated GOES imager output BTD values and MWPI values in the vicinity of downburst occurrence over western Oklahoma served as evidence of the presence of a well-developed mixed layer. Strong downbursts that were recorded by Oklahoma Mesonet stations between 0000 and 0100 UTC 27 August resulted from a combination of precipitation loading and sub-cloud evaporation of precipitation. These downbursts occurred in proximity to moderate to high microburst risk values as indicated in the 2200 UTC GOES microburst products.

A GOES imager microburst product (top) and a corresponding MWPI product (bottom) at 2200 UTC 26 August 2009, with the location of Oklahoma mesonet observations (i.e BESS, WEAT, etc.) of downburst wind gusts plotted on the MWPI image, are displayed in Figure 6. Both product images displayed convective storms developing along the cold front over western Oklahoma. Convective storm activity increased in coverage near the cold front during the following three hours. Downburst wind gusts between 41 and 56 knots were recorded by the Oklahoma Mesonet stations between 0000 and 0100 UTC 27 August. Significant downbursts recorded by the Oklahoma Mesonet during this event are noted in Table 2.

Figure 7, 0033 UTC Oklahoma City NEXRAD reflectivity overlying the 2200 UTC imager microburst product, displayed downburst-producing convective storms in progress west of Oklahoma City in a region of elevated microburst risk (orange shading). Also important to note the general increase in MWPI values from southwest (BESS) to northeast (MEDF) associated with a progression from hybrid to stronger wet type downbursts. Downbursts over western Oklahoma were predominantly "hybrid" type, while over north-central Oklahoma (MEDF, BREC), downbursts were "wet" type associated with heavy rainfall.

Figure 8 displays two RUC sounding profiles at 2200 UTC at Weatherford (top) and Medford (bottom), respectively, and shows contrasting downburst environments over Oklahoma. The Weatherford sounding, an "inverted-V" profile, indicates an overall deeper and drier mixed layer over western Oklahoma that favored the development of intense downdrafts due to the evaporation of precipitation in the sub-cloud layer. The Medford sounding, with a shallower, moister mixed layer and larger CAPE, indicated that precipitation loading was a significant factor in downdraft generation. Thus, this cold front downburst event demonstrates that favorable environments can vary over a relatively small geographic region. The stronger correlation between radar reflectivity and downburst wind gust speed for this event (.88) as compared to the 10 June event (.25) signified that precipitation loading was a more important factor in downburst magnitude, as typically expected in wet microburst environments.

## 4. Statistical Analysis and Discussion

Validation results for the 2007 to 2009 convective seasons have been completed for the MWPI and imager microburst products. GOES sounder-derived MWPI values have been compared to mesonet observations of downburst winds over Oklahoma and Texas for 168 events between June 2007 and September 2009. The correlation between MWPI values and measured wind gusts was determined to be .62 and was found to be statistically significant above the 99% confidence level, indicating that the correlation represents a physical relationship between MWPI values and downburst magnitude and is not an artifact of the sampling process. Comparison of GOES-11 imager microburst risk values (output brightness temperature difference (BTD) in degrees K) to measured downburst wind gusts for a total of 86 events in Idaho between July 2007 and September 2008 and in Oklahoma between June and September 2009 yielded a correlation of .51. The correlation between output BTD and measured wind gusts was determined to be statistically significant above the 99% confidence level, indicating a high confidence that the correlation represented a physical relationship between output BTD values and downburst magnitude. Also encouraging is the mean difference of 0.1 calculated between output BTD and wind gust speed. The small mean difference signifies the strength of the imager microburst product in distinguishing between the potential for severe convective wind gusts (> 50 knots) and wind gusts that would be considered operationally significant (35 to 50 knots).

Figure 9 shows scatterplots of MWPI values and GOES-11 imager output BTD values (K) versus observed downburst wind gust speed as recorded by mesonet stations in Oklahoma and Texas. The MWPI scatterplot identifies two clusters of values: MWPI values less than 50 that correspond to observed wind gusts of 35 to 50 knots, and MWPI values greater than 50 that correspond to observed wind gusts of greater than 50 knots. Similarly, the GOES-11 imager microburst risk scatterplot identifies two clusters. The dominant cluster contains output brightness temperature difference (BTD) values less than 50K that correspond to observed wind gusts between 35 and 50 knots. The scatterplots illustrate that both microburst products demonstrate effectiveness in distinguishing between severe and non-severe convective wind gust potential.

Further statistical analysis of a dataset built by comparing wind gust speeds recorded by Oklahoma Mesonet stations to adjacent MWPI values for 35 downburst events has yielded some favorable results, as displayed in Figures 10 and 11. Correlation was computed between key parameters in the downburst process, including temperature lapse rate (LR) and dewpoint depression difference (DDD) between two levels (670mb/850mb), CAPE, and radar reflectivity. The first important finding is a statistically significant negative correlation ($r=-.34$) between lapse rate and radar reflectivity. Similar to the findings of Srivastava (1985), for lapse rates greater than 8 K/km, downburst occurrence is nearly independent of radar reflectivity. For lapse rates less than 8 K/km, downburst occurrence was associated with high reflectivity (> 50 dBZ) storms. The majority of downbursts occurred in sub-cloud environments with lapse rates greater than 8.5 K/km. Adding the dewpoint depression difference to lapse rate yielded an even greater negative correlation ($r=-.42$) when compared to radar reflectivity. Finally, comparing the sum of LR and DDD (the former hybrid microburst index (HMI)) to CAPE resulted in the strongest negative correlation ($r=-.82$), with a confidence level above 99%. This emphasizes the complementary nature of the HMI and CAPE in generating a robust and physically meaningful MWPI value. This result also shows that CAPE can serve as an adequate proxy variable for precipitation loading (expressed as radar reflectivity) in the MWPI. The strong negative correlation, or negative functional relationship, between HMI and CAPE terms in the MWPI algorithm indicates that the MWPI should be effective in capturing both negative buoyancy generation and precipitation loading as downburst forcing mechanisms.

## 5. Summary and Conclusions

As documented in Pryor (2009) and in this paper, and proven by statistical analysis, the GOES sounder MWPI product has demonstrated capability in the assessment of wind gust potential over the southern Great Plains. In addition, a new multispectral GOES imager product has been developed and evaluated to assess downburst potential over the western United States. This microburst risk product image incorporates GOES-11 bands 3, 4, and 5 to sample the warm-season pre-convective environment and derive moisture stratification characteristics of the boundary layer that would be relevant in the analysis of microburst potential. Case studies and statistical analysis for downburst events that occurred during the 2007 to 2009 convective seasons demonstrated the effectiveness of the imager product with a significant correlation between risk values and microburst wind gust magnitude. The GOES-11 imager microburst product has been found to be effective in indicating the potential for dry microbursts. However, as noted by Caracena and Flueck (1988), the majority of microburst days during JAWS were characterized by environments intermediate between the dry and wet extremes (i.e. hybrid). As noted in Pryor (2009), the MWPI product is especially useful in the inference of the presence of intermediate or "hybrid" microburst environments, especially over the Great Plains region. The error curves in Figure 1 show that the ABI sounding offers a comparable performance to the GOES-12 sounding with a slight improvement to the temperature profile below the 850mb level. This finding underscores the importance of the 670 to 850mb temperature lapse rate in the assessment of downdraft instability and its inclusion in the MWPI algorithm.

## 6. References

Atkins, N.T., and R.M. Wakimoto, 1991: Wet microburst activity over the southeastern United States: Implications for forecasting. *Wea. Forecasting*, **6**, 470-482.

Brock, F. V., K. C. Crawford, R. L. Elliott, G. W. Cuperus, S. J. Stadler, H. L. Johnson and M. D. Eilts, 1995: The Oklahoma Mesonet: A technical overview. *Journal of Atmospheric and Oceanic Technology*, **12**, 5-19.

Caracena, F., and J.A. Flueck, 1988: Classifying and forecasting microburst activity in the Denver area. *J. Aircraft*, **25**, 525-530.

Fujita, T.T., 1971: Proposed characterization of tornadoes and hurricanes by area and intensity. SMRP Research Paper 91, University of Chicago, 42 pp.

Fujita, T.T., and H.R. Byers, 1977: Spearhead echo and downburst in the crash of an airliner. *Mon. Wea. Rev.*, **105**, 129–146.

Pryor, K.L., and G.P. Ellrod, 2004: WMSI - A new index for forecasting wet microburst severity. *National Weather Association Electronic Journal of Operational Meteorology*, 2004-EJ3.

Pryor, K.L., 2008: An initial assessment of the GOES Microburst Windspeed Potential Index. Preprints, *5th GOES Users' Conf.*, New Orleans, LA, Amer. Meteor. Soc.

Pryor, K.L., 2009: Microburst windspeed potential assessment: progress and developments. Preprints, 16th Conf. on Satellite Meteorology and Oceanography, Phoenix, AZ, Amer. Meteor. Soc.

Przybylinski, R.W., 1995: The bow echo. Observations, numerical simulations, and severe weather detection methods. *Wea. Forecasting*, **10**, 203-218.

Schaefer, J.T., 1986: The dryline. In Mesoscale Meteorology and Forecasting. P.S. Ray (Ed.), American Meteorological Society, Boston, 549-572.

Schmit, T.J., J. Li, J.J. Gurka, M.D. Golberg, K.J. Schrab, J. Li, and W. F. Feltz, 2008: The GOES-R

Advanced Baseline Imager and the Continuation of Current Sounder Products. Journal of Applied Meteorology and Climatology, **47**, 2696–2711.

Schroeder, J.L., W.S. Burgett, K.B. Haynie, I. Sonmez, G.D. Skwira, A.L. Doggett, and J.W. Lipe, 2005: The West Texas Mesonet: A technical overview. *Journal of Atmospheric and Oceanic Technology*, **22**, 211-222.

Srivastava, R.C., 1985: A simple model of evaporatively driven downdraft: Application to microburst downdraft. *J. Atmos. Sci.*, **42**, 1004-1023.

Wakimoto, R.M., 1985: Forecasting dry microburst activity over the high plains. *Mon. Wea. Rev.*, **113**, 1131-1143.

**Acknowledgements**

The author thanks Mr. Derek Arndt (Oklahoma Climatological Survey)/Oklahoma Mesonet, and Dr. John Schroeder (Texas Tech University)/West Texas Mesonet for the surface weather observation data used in this research effort. The author also thanks Jaime Daniels (NESDIS) for providing GOES sounding retrievals displayed in this paper.

**Correlation:**

| | | | |
|---|---|---|---|
| MBR to measured wind: | 0.64 | Mean LFC: | 674 |
| Reflectivity to measured wind: | 0.25 | Mean Wind Speed: | 45.8 |
| Mean Diff. (MBR, Wind Speed): | 1.20 | Mean MBR: | 47 |

| Date | Time (UTC) | Measured Wind Speed kt | Location | GOES-11 MBR K | Radar dBZ |
|---|---|---|---|---|---|
| **10-Jun-09** | 2155 | 50 | Hereford TX | 49 | 50 |
| | 2205 | 38 | Dimmitt TX | 43 | 40 |
| | 2255 | 53 | Dora NM | 51 | 40 |
| | 2320 | 48 | Kenton OK | 44 | 65 |
| | 2355 | 40 | Boise City OK | 48 | 45 |

Table 1. Performance statistics for 10 June 2009 downburst event.

**Correlation:**

| | | | |
|---|---|---|---|
| MWPI to measured wind: | 0.42 | Mean LFC: | 750 |
| MBR to measured wind: | 0.55 | Mean Wind Speed: | 44.88 |
| Reflectivity to measured wind: | 0.88 | Mean MBR: | 46.75 |
| | | Mean Diff. (MBR, Wind): | 1.88 |

| Date | Time (UTC) | Measured Wind Speed kt | Location | GOES-12 MWPI | GOES-11 MBR K | Radar dBZ |
|---|---|---|---|---|---|---|
| **27-Aug-09** | 0005 | 50 | Bessie OK | 30 | 51 | 50 |
| | 0020 | 43 | Kingfisher OK | 46 | 50 | 44 |
| | 0030 | 41 | Weatherford OK | 34 | 48 | 43 |
| | 0040 | 50 | El Reno OK | 56 | 49 | 49 |
| | 0055 | 56 | Medford OK | 57 | 48 | 54 |
| | 0110 | 40 | Breckinridge OK | 57 | 48 | 47 |
| | 0240 | 36 | Kenton OK | 32 | 39 | 33 |
| | 0310 | 43 | Boise City OK | 37 | 41 | 41 |

Table 2. Performance statistics for 27 August downburst event.

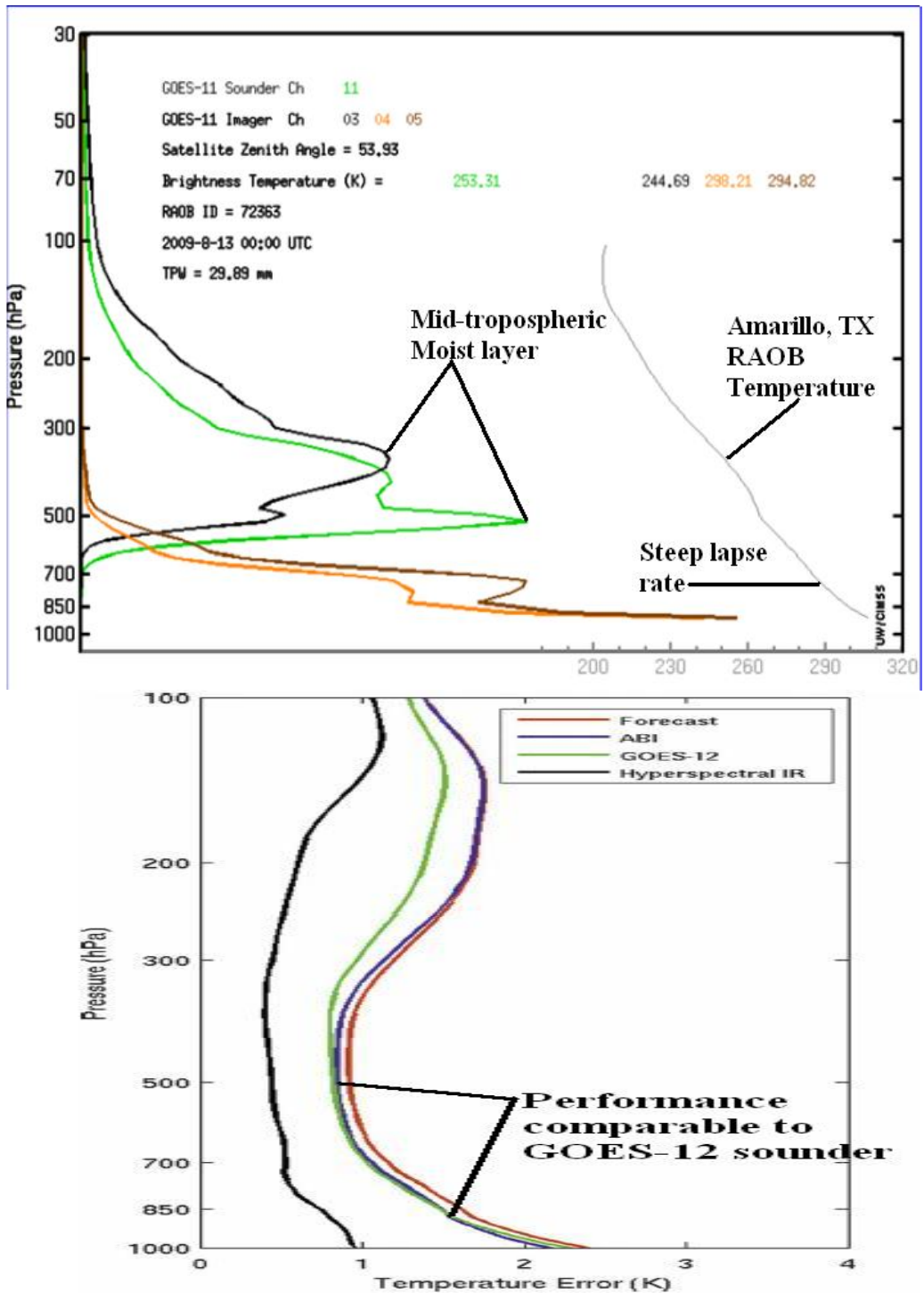

Figure 1. GOES-11 imager transmittance weighting functions compared to the radiosonde observation (RAOB) profile over Amarillo, Texas at 0000 UTC 13 August 2009 with temperature curve displayed (top) and temperature error curves for NWP forecast, ABI, GOES-12, and Hyperspectral IR soundings (courtesy Schmit et al. (2008), bottom).

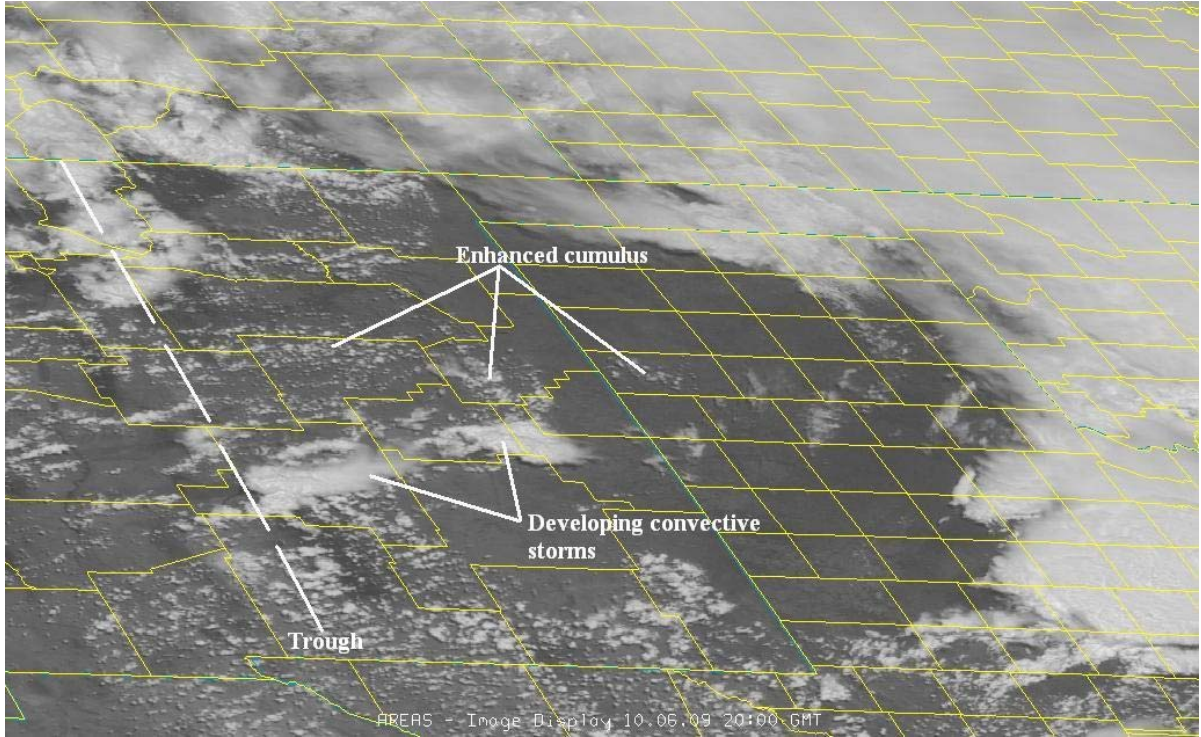
Figure 2.  GOES-11 visible image at 2000 UTC 10 June 2009.

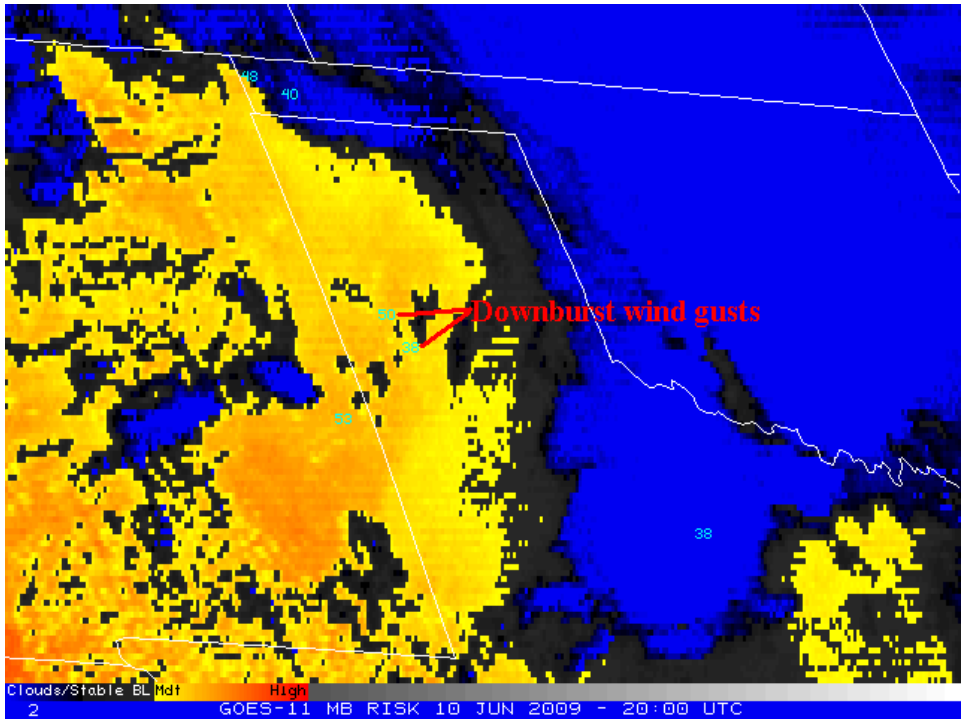

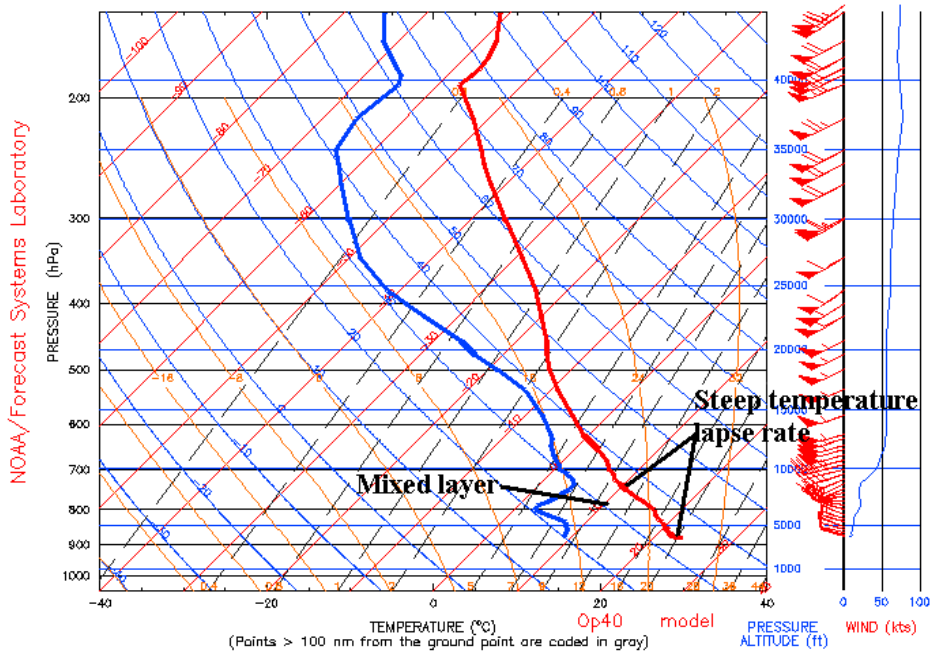

Figure 3. GOES-11 imager microburst risk product at 2000 UTC 10 June 2009 (top) with mesonet observations of downburst wind gusts plotted on the image, and a corresponding RUC sounding at Hereford, Texas.

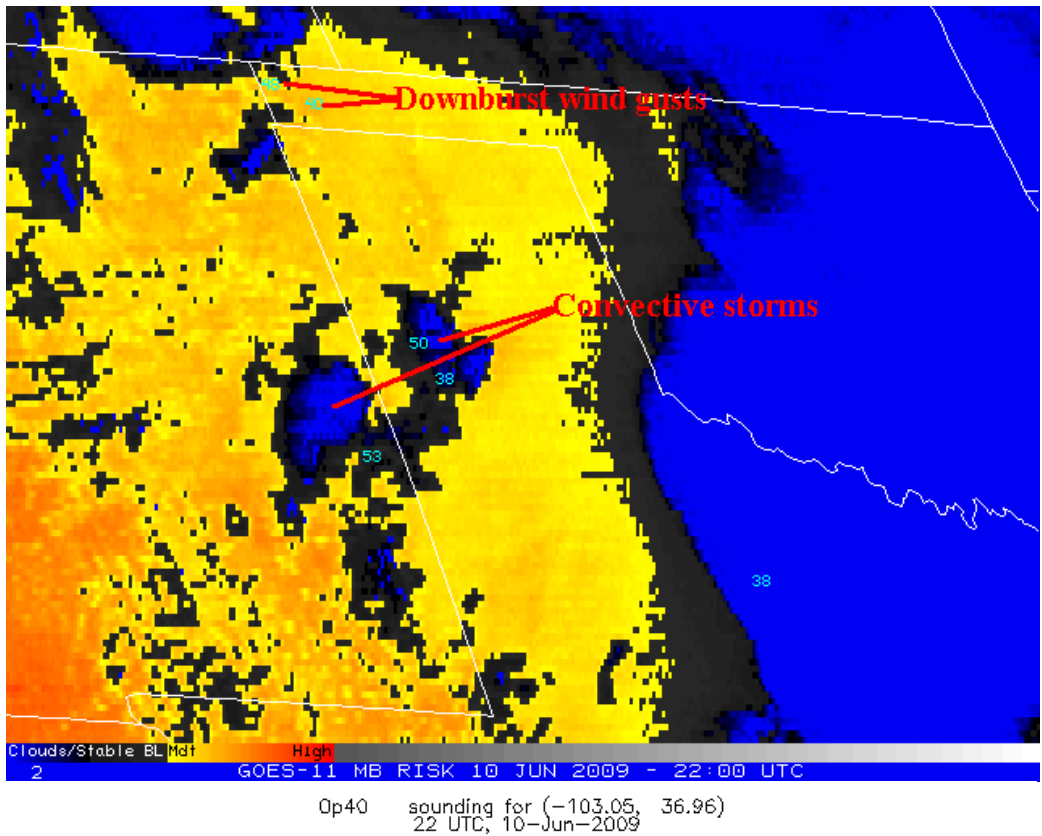

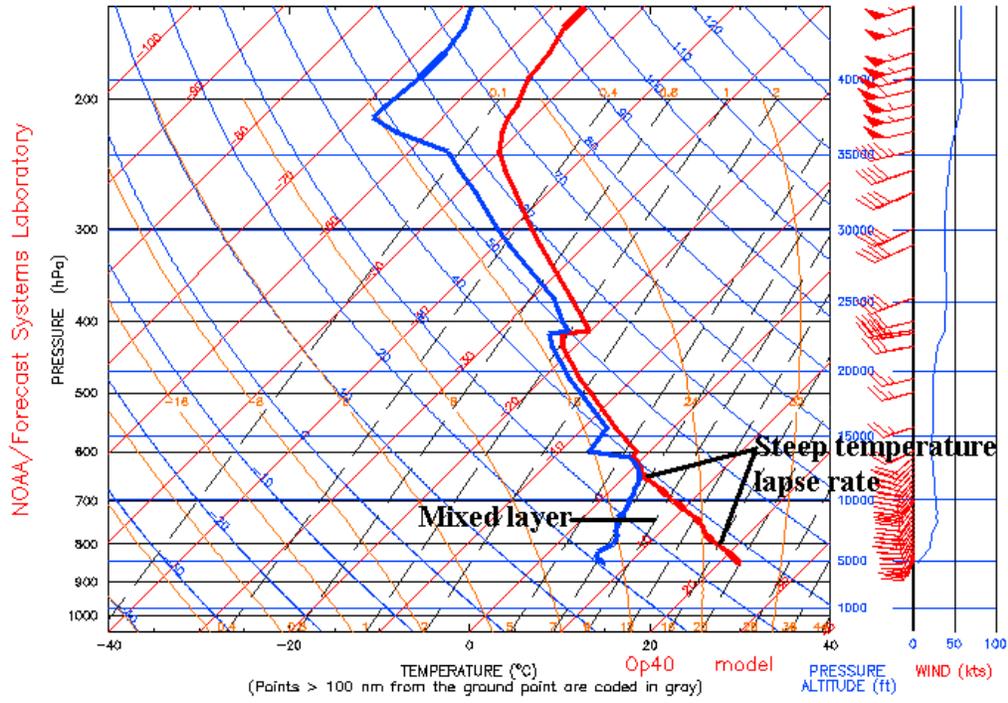

Figure 4. GOES-11 imager microburst risk product at 2200 UTC 10 June 2009 (top), with mesonet observations of downburst wind gusts plotted on the image, and a corresponding RUC sounding at Kenton, Oklahoma.

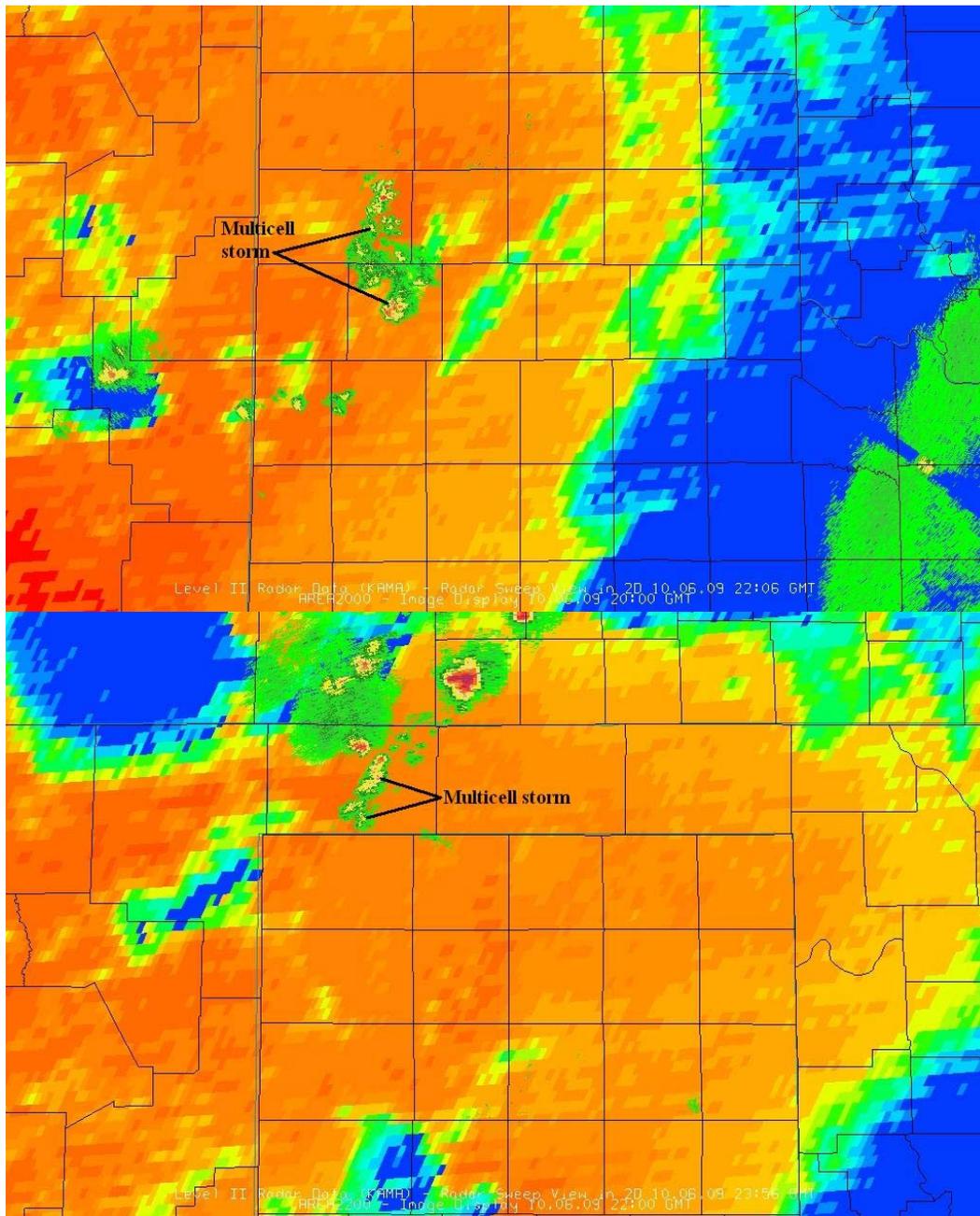

Figure 5. GOES-11 imager microburst risk products at 2000 UTC (top) and 2200 UTC (bottom) 10 June 2009 with overlying radar reflectivity imagery from Amarillo NEXRAD. These images are visualized by McIDAS-V software.

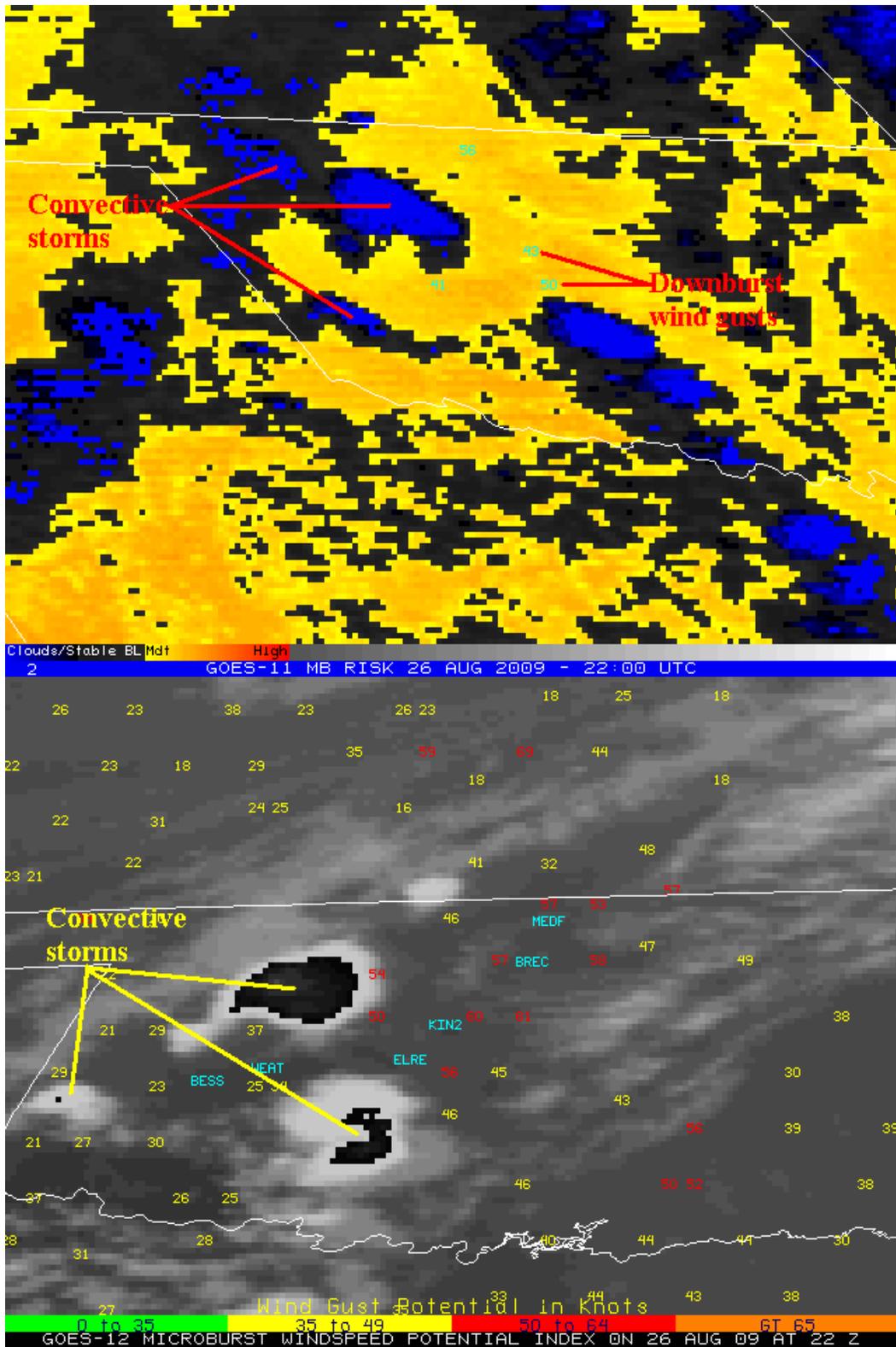

Figure 6. GOES-11 imager microburst risk product (top) and GOES-12 MWPI product (bottom) at 2200 UTC 26 August 2009, with mesonet observations of downburst wind gusts plotted on the image.

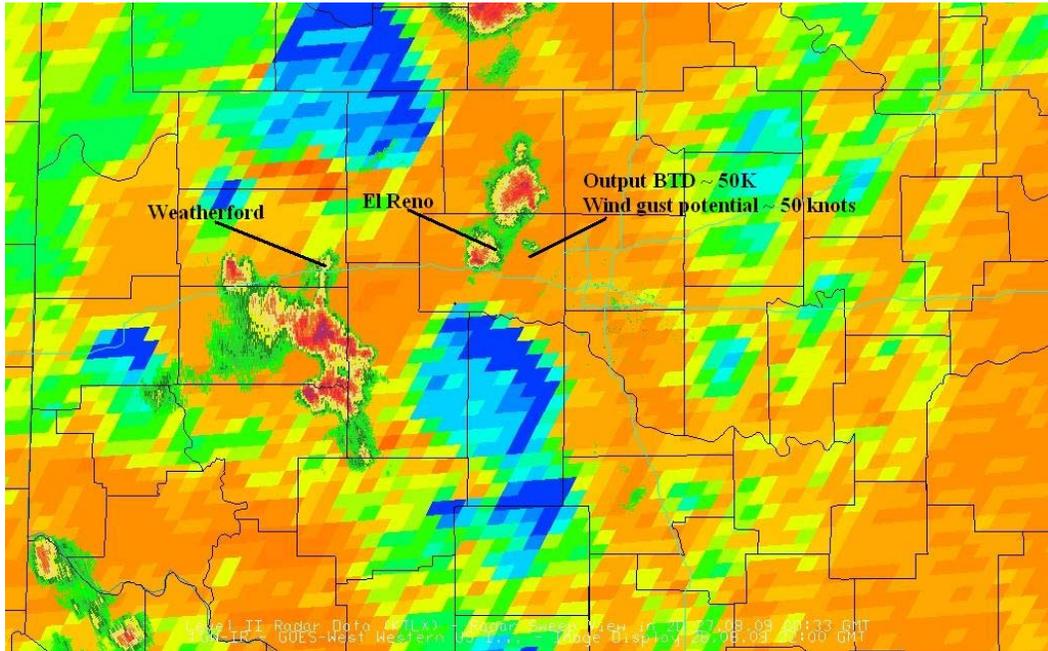

Figure 7.  0033 UTC Oklahoma City NEXRAD reflectivity overlying the 2200 UTC 26 August 2009 imager microburst product.

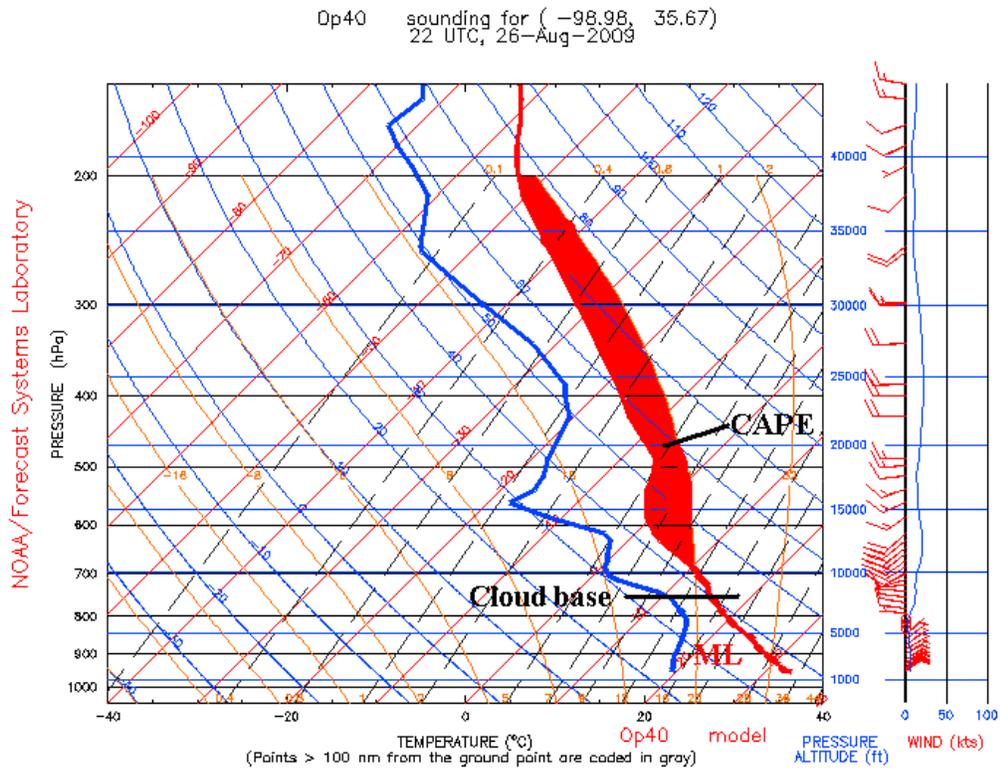
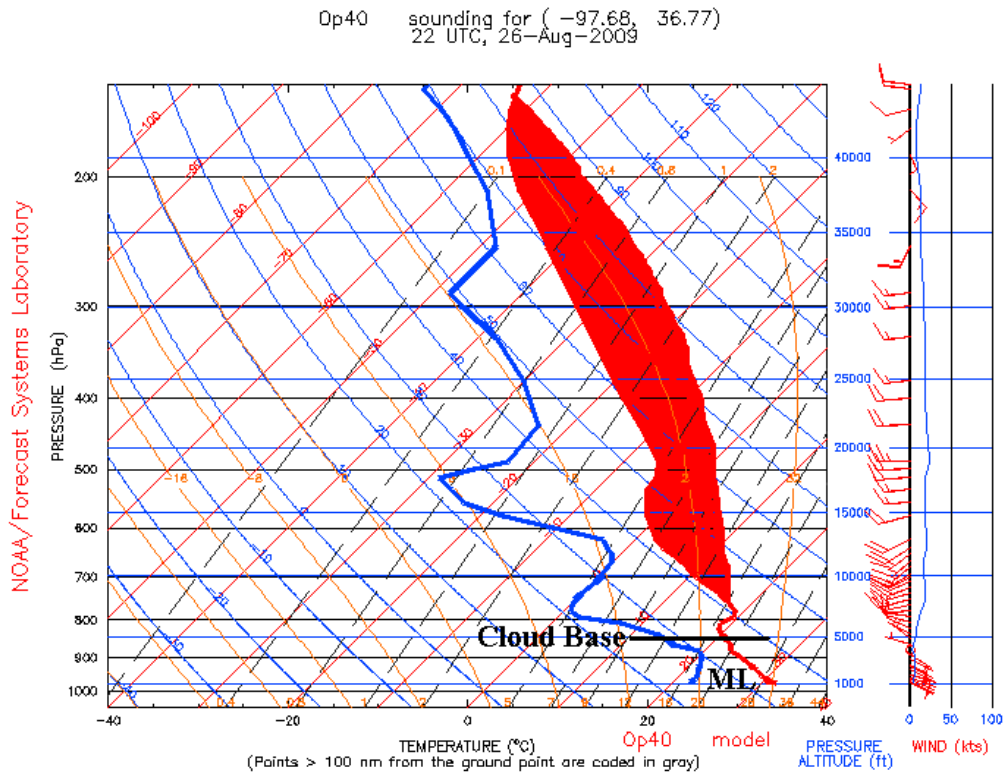

Figure 8. RUC sounding profiles at 2200 UTC 26 August 2009 at Weatherford (top) and Medford (bottom), Oklahoma mesonet stations.

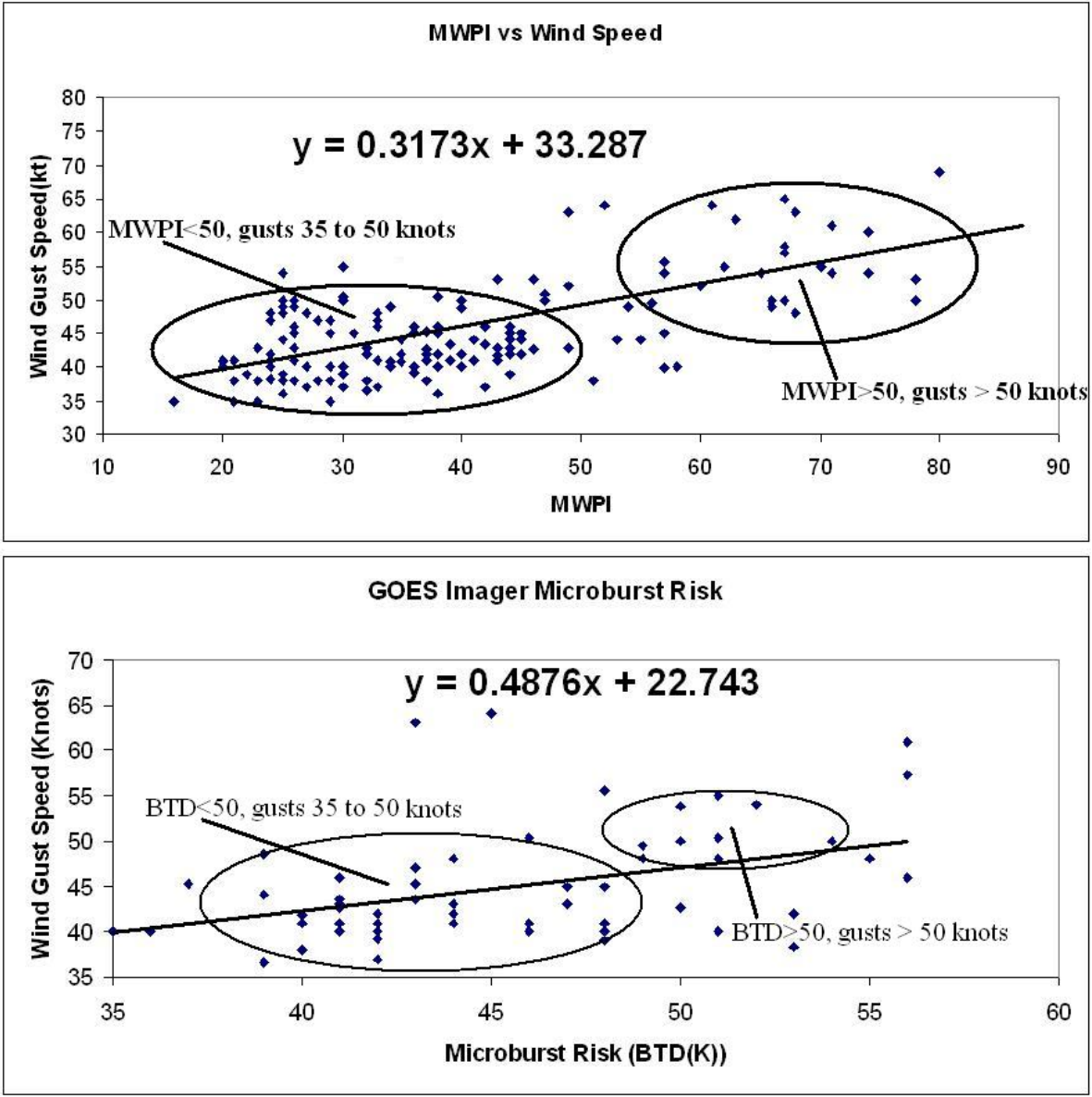

Figure 9. Scatterplots of MWPI values (top) and GOES-11 imager output BTD values (K)(bottom) versus observed downburst wind gust speed as recorded by mesonet stations in Oklahoma and Texas.

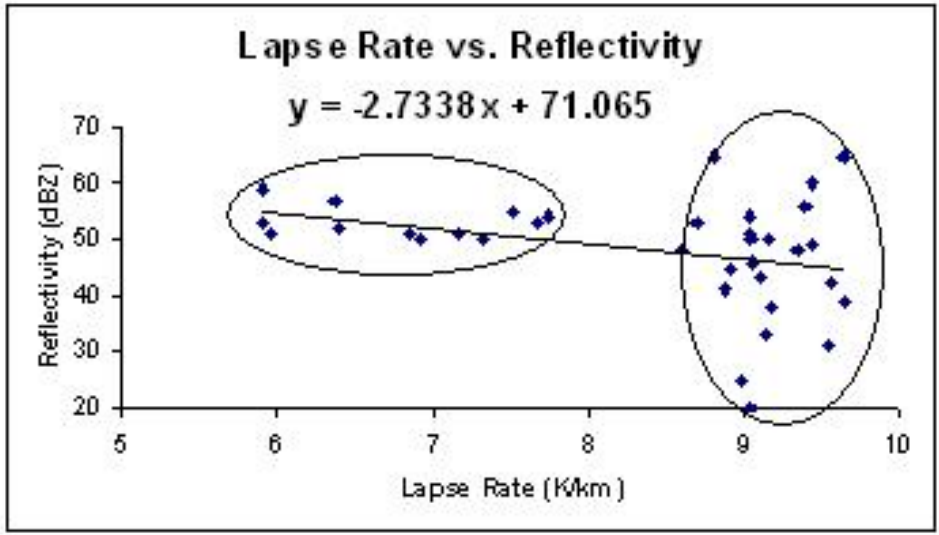

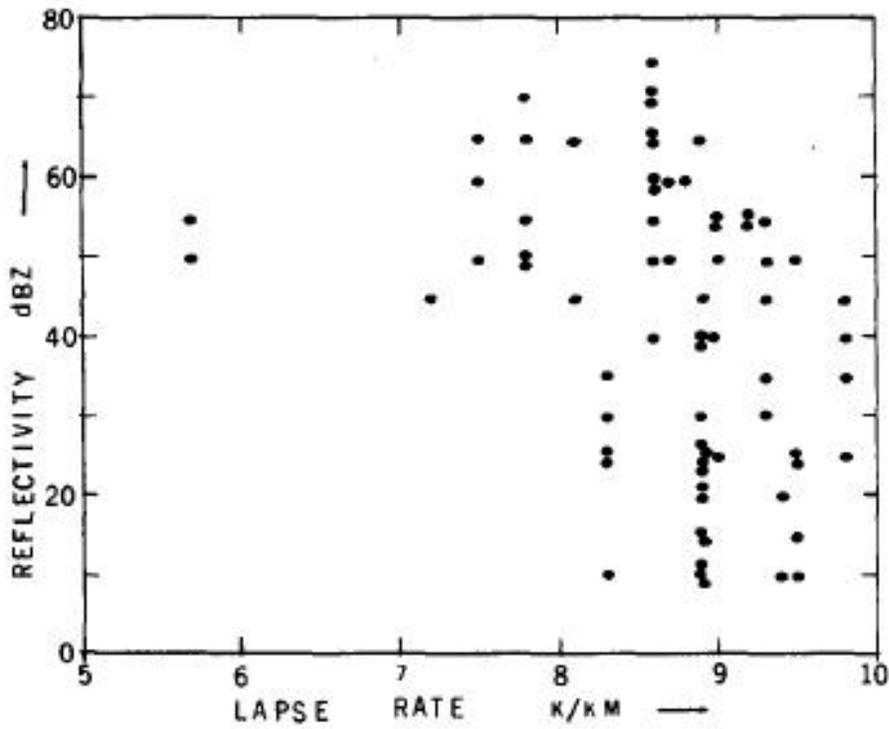

Figure 10. Scatterplot of lapse rate versus radar reflectivity for 35 downburst events over Oklahoma during the 2009 convective season (top) compared to scatterplot for 186 microburst events during the 1982 JAWS project (courtesy Srivastava 1985).

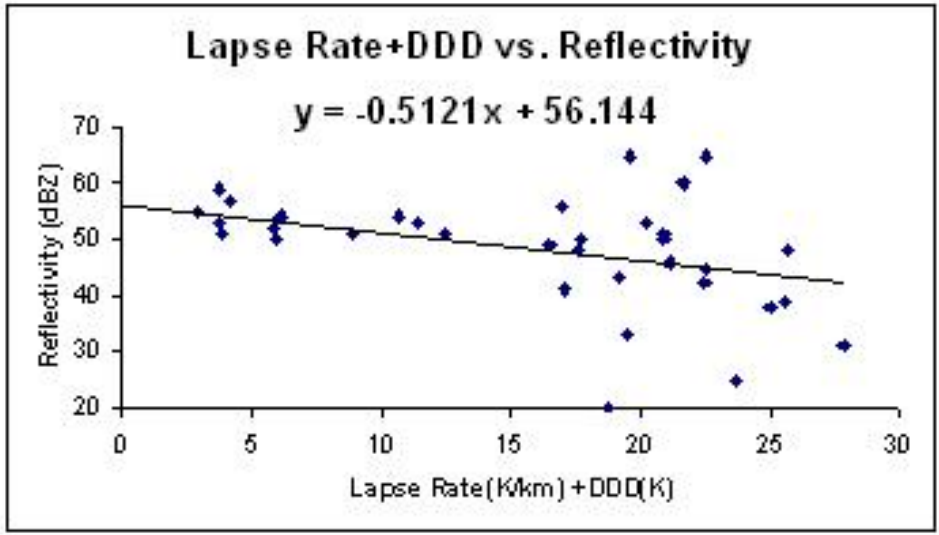

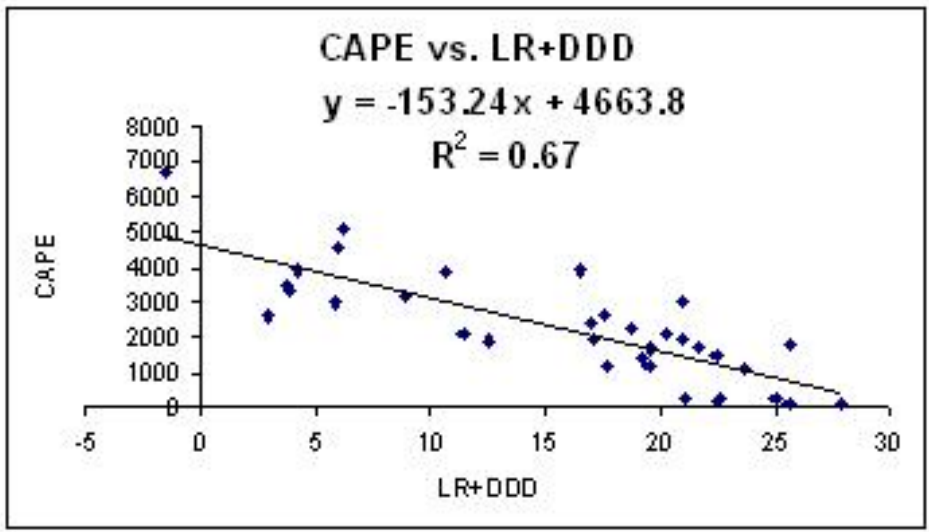

Figure 11. Scatterplot of the sum of lapse rate and DDD versus radar reflectivity (top) compared to scatterplot of the sum of lapse rate and DDD versus CAPE (bottom) for 35 downburst events over Oklahoma during the 2009 convective season.